\documentclass[conference]{IEEEtran}
\IEEEoverridecommandlockouts
\usepackage{cite}
\usepackage{amsmath,amssymb,amsfonts}
\usepackage{algorithmic}
\usepackage{graphicx}
\usepackage{textcomp}
\usepackage{xcolor}
\usepackage{balance}
\usepackage{hyperref}
\usepackage{todonotes}
\usepackage{booktabs}
\usepackage{siunitx}
\usepackage{multirow}
\usepackage{subcaption}
\usepackage{tcolorbox}

\def\BibTeX{{\rm B\kern-.05em{\sc i\kern-.025em b}\kern-.08em
    T\kern-.1667em\lower.7ex\hbox{E}\kern-.125emX}}
\begin{document}

\title{
HTML Structure Exploration in 3D Software Cities
}

\author{\IEEEauthorblockN{Malte Hansen}
\IEEEauthorblockA{\textit{Department of Computer Science} \\
\textit{Kiel University}\\
Kiel, Germany \\
malte.hansen@email.uni-kiel.de}
\and
\IEEEauthorblockN{David Moreno-Lumbreras}
\IEEEauthorblockA{\textit{Escuela de Ingeniería de Fuenlabrada} \\
\textit{Universidad Rey Juan Carlos }\\
Fuenlabrada, Spain \\
david.morenolu@urjc.es}
\and
\IEEEauthorblockN{Wilhelm Hasselbring}
\IEEEauthorblockA{\textit{Department of Computer Science} \\
\textit{Kiel University}\\
Kiel, Germany \\
hasselbring@email.uni-kiel.de}
}

\maketitle

\begin{abstract}
Software visualization, which uses data from dynamic program analysis, can help to explore and understand the behavior of software systems.
It is common that large software systems offer a web interface for user interaction.
Usually, available web interfaces are not regarded in software visualization tools.
This paper introduces additions to the web-based live tracing software visualization tool ExplorViz:
We add an embedded web view for instrumented applications in the 3D visualization to ease interaction with the given applications and enable the exploration of the thereby displayed HTML content.
Namely, the Document Object Model (DOM) is visualized via a three-dimensional representation of the HTML structure in same-origin contexts.

Our visualization approach is evaluated in a preliminary user study.
The study results give insights into the potential use cases, benefits, and shortcomings of our implemented approach.
Based on our study results, we propose directions for further research to support the visual exploration of web interfaces and explore use cases for the combined visualization of software cities and HTML structure.

Video URL: https://youtu.be/wBWKlbvzOOE
\end{abstract}

\begin{IEEEkeywords}
software visualization, city metaphor, web, 3D, html, program comprehension
\end{IEEEkeywords}

\section{Introduction}\label{sec:introduction}
Data visualization plays a central role in understanding complex structures in software systems~\cite{koschke2003software}. Traditionally, 2D techniques such as heatmaps and metric charts have been employed to assist developers in comprehending large codebases~\cite{Ball1996, Lanza2007}. These approaches are integrated into modern IDEs and platforms like GitHub or Visual Studio Code to support software maintenance and evolution~\cite{bacchelli2013expectations, thongtanunam2017review, yu2015wait}. They are useful for identifying hotspots~\cite{Meyer2008} and detecting low-quality code regions, but the visualization of hierarchical, multidimensional, or temporally evolving structures like the Document Object Model (DOM) poses a challenge.

To address these limitations, several efforts have explored the use of three-dimensional representations.
The city metaphor, in particular, has been successfully used to convey the structure and behavior of software systems~\cite{knight1999, codecity}.
Our tool ExplorViz is prominent in this space, offering near real-time 3D visualizations of software execution traces~\cite{fittkau2017, hasselbring2020}. 
It enables developers to explore runtime behavior and architectural components in a spatial layout.

Inspecting the runtime behavior of a software system usually means interacting with its web interface in a separate browser tab.
This method requires knowledge of the web address on which the web front end is hosted, and it introduces context switches when leaving the 3D scene of the software visualization.
These context switches become even more prevalent when analyzing the DOM of the software under inspection within the browser.

To address the given issue, this paper introduces two conceptual additions to 3D software cities, which are presented through a prototypical implementation in ExplorViz.
First, an embedded web view for instrumented applications is integrated in the 3D visualization to facilitate the interaction and generation of trace data with the given applications and avoid leaving the 3D scene.
Secondly, we enable the exploration of the thereby displayed HTML content in same-origin contexts.
Namely, the Document Object Model (DOM) is visualized via a three-dimensional representation of the HTML structure.

The remainder of this paper is structured as follows.
In Section \ref{sec:related-work}, we discuss related work in the field of visualizing structured HTML.
In Section \ref{sec:web-view}, we describe the integration of HTML content in iframes into the 3D software landscape by means of embedded web views.
Section \ref{sec:html-visualization} focuses on the interactive 3D visualization of HTML that is displayed in the embedded web view.
A preliminary evaluation and the feedback collected are presented in Section \ref{sec:evaluation}.
To summarize our approach, Section \ref{sec:conclusion} concludes the paper and outlines the potential for future research.

\section{Related Work}\label{sec:related-work}
Our approach builds on top of the existing live trace visualization tool ExplorViz~\cite{fittkau2017, hasselbring2020}.
The tool processes data from dynamic software analysis to visualize the structure of software systems as cities.
ExplorViz is developed as an open source project.\footnote{\url{https://github.com/ExplorViz}}

In this line of work, other approaches have been extended to the domain of HTML and DOM for spatial three-dimensional visualizations.
BabiaXR introduced a 3D representation of the DOM structure, where HTML elements are displayed as buildings within a city metaphor~\cite{Moreno-Lumbreras_2024}. That approach included real-time synchronization with code editors like Visual Studio Code and was particularly aimed at improving comprehension of complex web interfaces through interactive overlays and structured layouts.
Our approach for HTML visualization (see Section \ref{sec:html-visualization}) is a re-implementation and extension of the HTML visualizer used in BabiaXR.
Other tools have also explored 3D visualization of the DOM.
Mozilla's Tilt~\cite{tilt} and Microsoft's 3D View~\cite{3dview} provide representations of HTML documents in three dimensions directly build into browsers.
However, these tools are limited to static snapshots of the DOM and do not support integration with dynamic analysis or development environments.
They lack features for runtime instrumentation or synchronized exploration alongside application logic.

The approach introduced in this paper builds upon the concept of three-dimensional DOMs by integrating DOM explorations directly into the ExplorViz platform. 
Unlike previous tools, the DOM visualization is embedded within a live software city, enabling joint inspection of interface and execution behavior.
This represents a step forward toward full-stack visualization for modern web-based software systems.

\section{Embedded Web View}\label{sec:web-view}
\begin{figure}[htbp]
	\includegraphics[width=\textwidth/2]{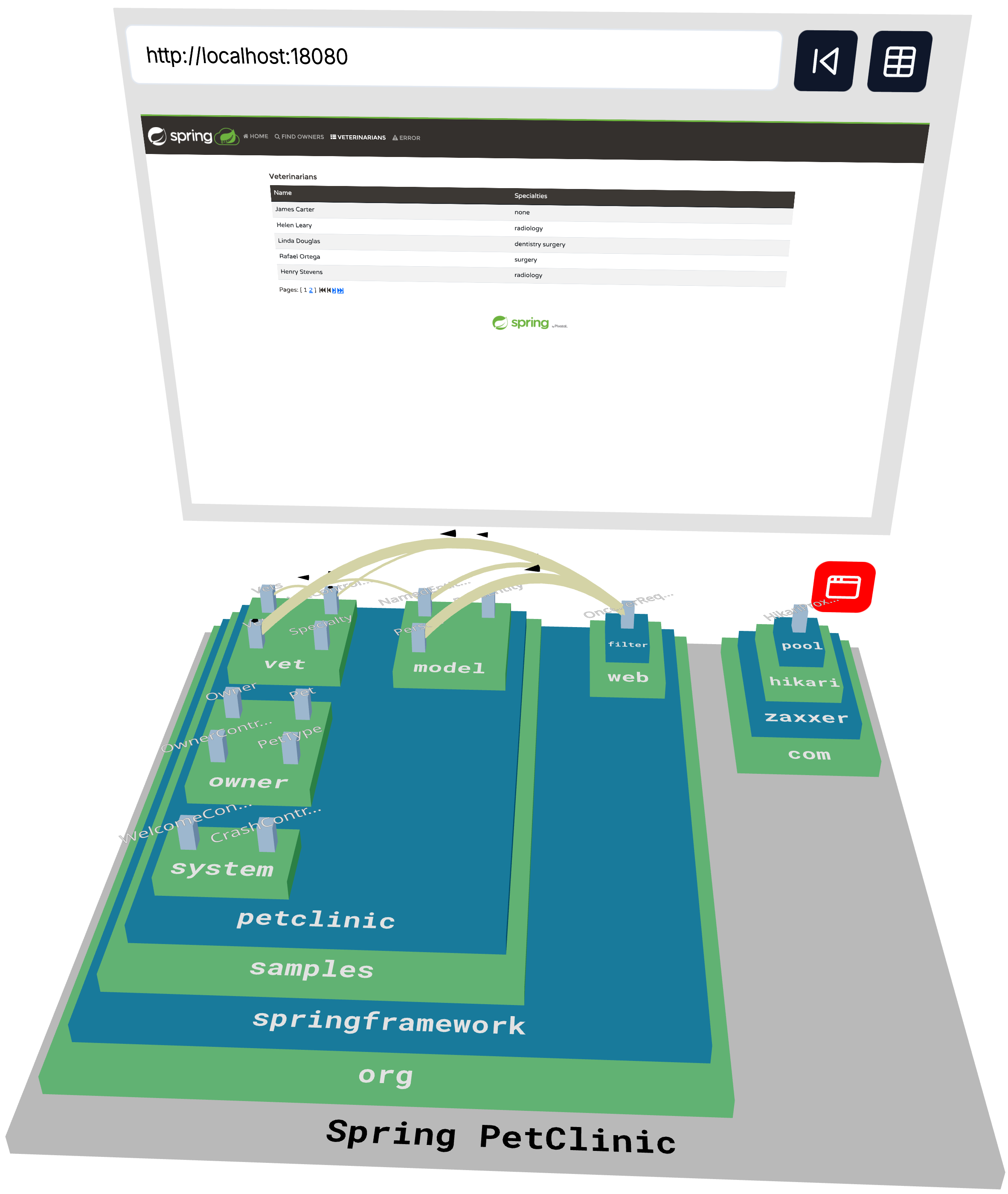}
	\caption{Software city of the dynamically instrumented PetClinic with opened embedded browser. User input on the embedded iframe leads to the generation of trace data which is reflected in the software city.}
	\label{fig:petclinic}
\end{figure}

To explore the behavior of applications with tools such as ExplorViz, a user may want to interact with the web interface of the given application.
Using the React Drei HTML component,\footnote{\url{https://drei.docs.pmnd.rs/misc/html}} we are able to render HTML content in an iframe as part of the 3D scene.
This is augmented with an input field for URLs and buttons to go back to the default URL or open the HTML visualization (see Section \ref{sec:html-visualization}) respectively.
An example for the resulting visualization is presented in Figure \ref{fig:petclinic} that displays the Spring PetClinc with the corresponding web interface embedded in the 3D scene.
In this paper, we refer to the iframe with its user interface elements as embedded web view, as it adds an additional view displaying web content into the 3D software city.
The embedded web view can be toggled via a button in the back of a visualized application.
For our implementation, the web view navigates to a fixed default URL after it was opened.
However, the address can easily be configured for each individual application and even extracted from the available monitoring data.
Thereby, when displaying a large and distributed software landscape, multiple embedded web views may be opened at once, each displaying the web interface that corresponds to the given application. 

The embedded web view can display any content that can be displayed in regular iframes, including locally accessible HTML documents.
However, the inclusion of content in iframes is limited by a browser's security policy.
In addition, websites can set HTTP headers to instruct browsers not to load the given website as an iframe at all.\footnote{\url{https://developer.mozilla.org/en-US/docs/Web/HTTP/Reference/Headers/X-Frame-Options}}
Since software visualization is usually used to inspect software under development and therefore under the control of the respective user, this is a negligible limitation.

\section{HTML Visualization}\label{sec:html-visualization}
\begin{figure*}
  \centering
  \includegraphics[width=\textwidth]{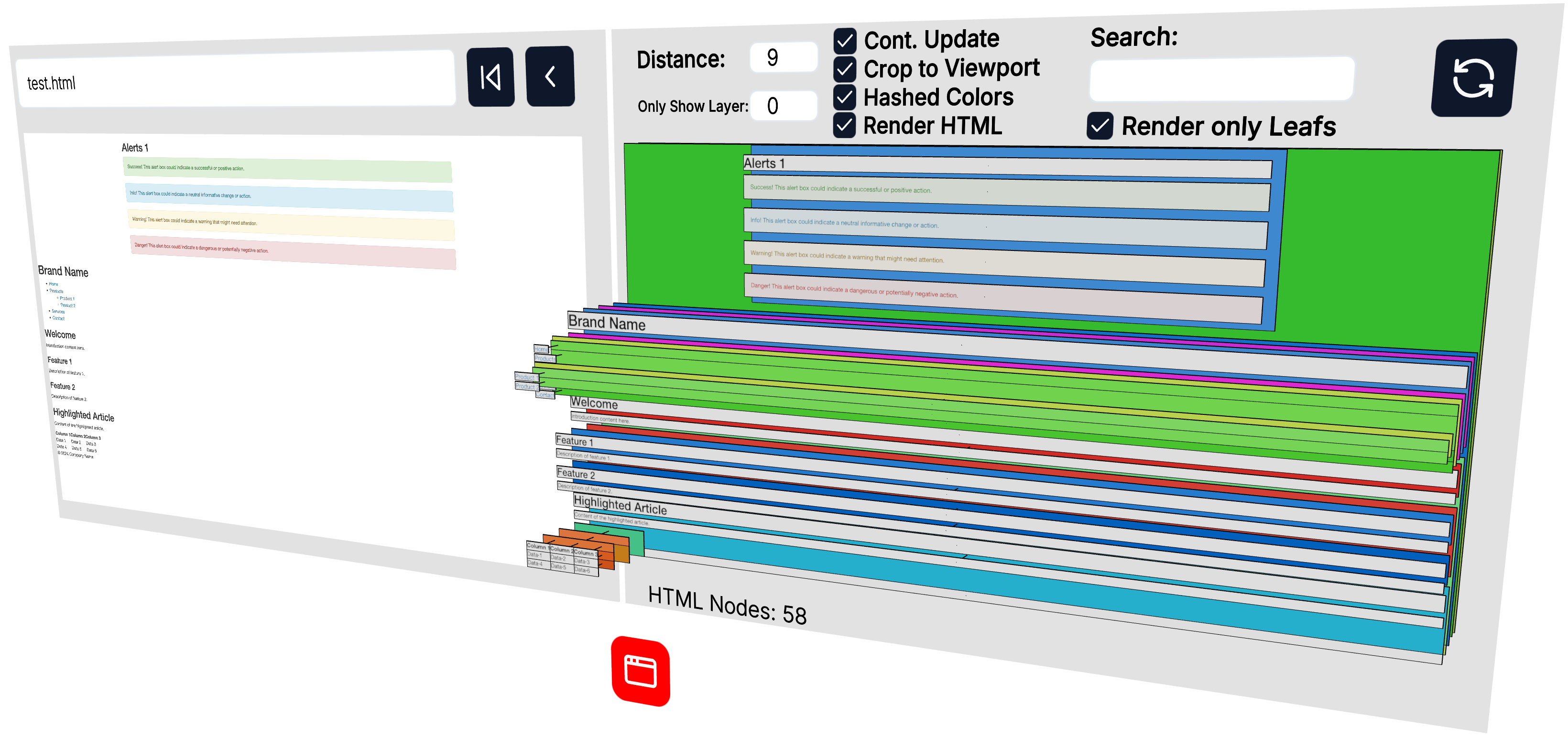}
  \caption{A locally accessible HTML document (compare to~\cite{Moreno-Lumbreras_2024}) is displayed in the embedded web view (left) and augmented with the visualization of its structure (right). Leaf nodes of the DOM are given a rendered texture while other nodes are colored according to their similarity.}
  \label{fig:test-html}
\end{figure*}

The web view can be visually extended with a view to interactively analyze the HTML structure given by the underlying DOM.
Regarding its concept and implementation, this process is closely related to the visualization in BabiaXR~\cite{Moreno-Lumbreras_2024}.
The tree structure of the iframe's DOM is processed and mapped to colored 3D boxes in the visualization (see Figure \ref{fig:test-html}).
By default, every change in the observed DOM leads to an update in the visualization.
Alternatively, e.g. for web pages that frequently update their DOM, continuous updating may be disabled, and a manual refresh of the visualization can be triggered instead.

Each visual layer contains nodes at a certain depth of the DOM's tree.
These nodes are represented as colored boxes.
Black lines visually connect boxes to their parent element in the DOM.
The distance between the layers and the layer that should be displayed can be changed via the input fields.
These numerical inputs can also be updated incrementally by using the horizontal scroll wheel of a mouse.

By default, all elements on a specific layer have the same color.
As an alternative, the boxes can be colored according to a hash function which primarily takes into account the tag name of the underlying element (see Figure \ref{fig:test-html}).
The package html-to-image\footnote{\url{https://github.com/sasithahtl/htmltoimage}} enables us to take screenshots of the displayed web page.
These images can be mapped to leaf nodes or projected onto all 3D boxes in the viewport.

It is also possible to display the extension of elements beyond the viewport, taking into account scrollable or otherwise hidden content (see Figure \ref{fig:study-viewport}).
In addition to filtering by layer and applying viewport cropping, a user can further narrow down the selection by means of a text search.
This text search takes into account the opening and closing HTML tags, including all attributes and classes as well as textual content inside the HTML elements.
Exemplary, searching for "\textless img" would show all images with the corresponding tag name.
The number of currently visualized HTML elements, taking into account all filter options, is displayed at the bottom of the HTML visualization.

Regarding interactivity, the boxes can be hovered to display the same text that is also used for the textual search.
Clicking on a box enlarges the box, shows a popover with the textual representation of the HTML, and hides the boxes of all elements in the DOM that do not belong to the subtree of the clicked box's HTML element.
A double click on a box sends a click event to the corresponding HTML element, enabling basic navigation in the HTML visualization and making it accessible which elements have registered click listeners.

\textbf{Limitations}
There are many known attacks which take advantage of browsers loading foreign content, possibly exposing sensitive user data~\cite{browser-security}.
Therefore, browsers only permit programmatically analyzing iframes which have the same source, i.e. the same domain and same port, as the current webpage.
For subdomains, this restriction may be relaxed if both pages include the same base domain as an attribute in the document model.
HTML that is processed and received from internal back-end services can be analyzed.

\section{Preliminary Evaluation}\label{sec:evaluation}
To obtain early insights and collect feedback on the applicability of our embedded browser and the HTML visualization, we conducted a small user study.
This preliminary study was motivated by two research questions:

\begin{itemize}
   \item \textbf{R1} Is the embedded browser a viable alternative to a separate browser tab to interact with a software under inspection?
   \item \textbf{R2} Is the HTML visualization a viable addition to established web tools to explore HTML?
\end{itemize}

The data obtained for our evaluation, a demonstration video, and software packages are publicly available~\cite{hansen_2025_15518604}.

\textbf{Participants}
We asked people to evaluate our approach, who actively worked as students or researchers in the development of ExplorViz.
This group of people is already familiar with web development and ExplorViz, such that they can focus on the evaluation of our newly added approach to HTML structure visualization.
Participants in this study did not contribute to the conception or development of the approach presented in this article.
Six people volunteered for our study.

\begin{figure}[htbp]
	\includegraphics[width=\textwidth/2]{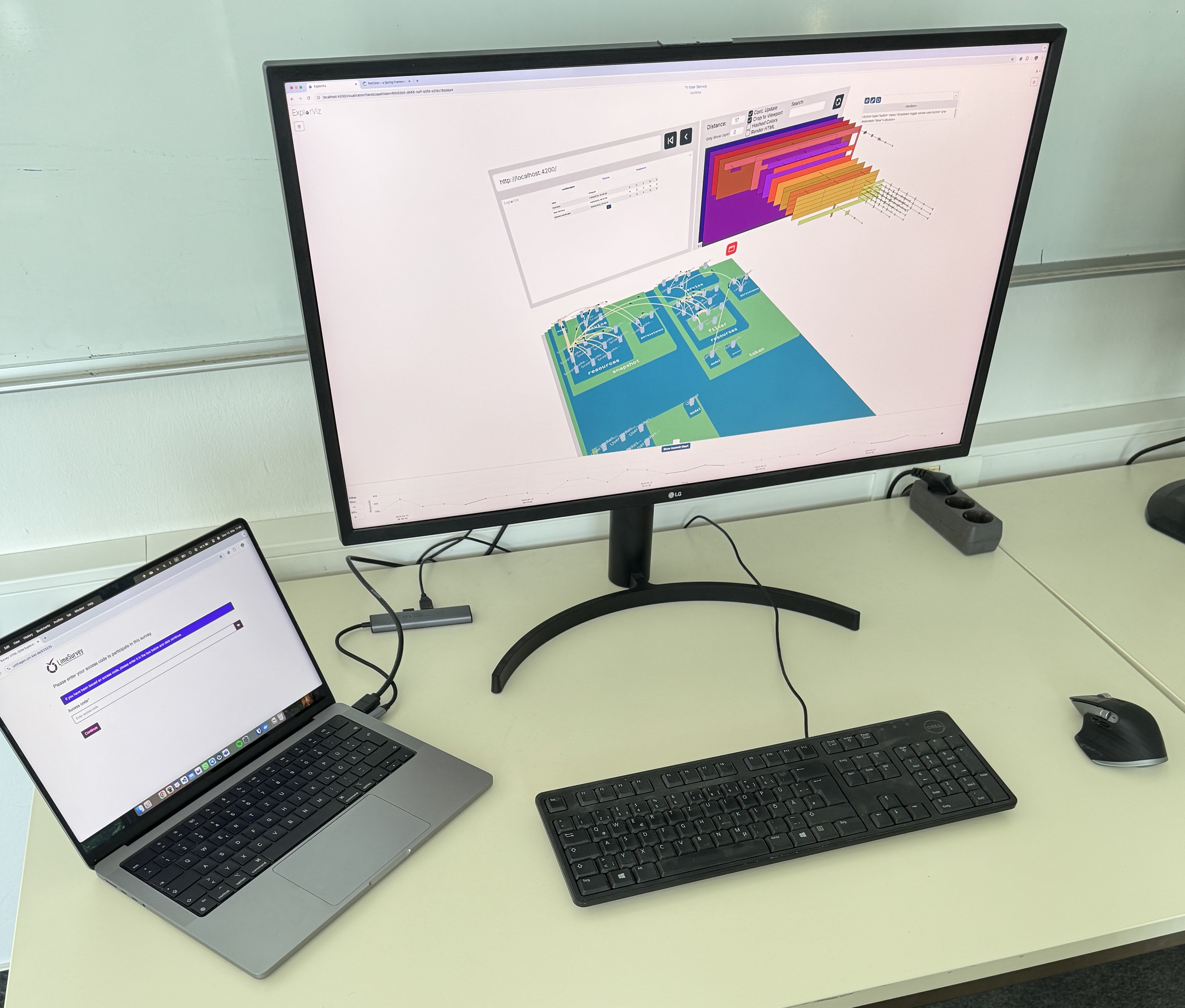}
	\caption{Study setup where the internal laptop monitor displays the digital survey and the external monitor displays the visualization of the ExplorViz user service and the HTML visualization of the landscape selection screen in  ExplorViz.}
	\label{fig:study-setup}
\end{figure}

\textbf{Hardware Setup}
The hardware setup for the study in a lab environment included an Apple MacBook Pro with M1 Pro chipset, a 32 inch Ultra HD monitor, and a mouse which comes with a second scroll wheel for horizontal scrolling.
A large monitor with a high resolution was chosen, to ensure that rendered content in the embedded browser is readable while still being able to inspect the software city.
The internal monitor of the laptop was used to display the survey and tasks to the participants.
The second scroll wheel of the mouse enabled participants to easily change the distance between layers or restrict the view to specific layers.
The setup of the study is depicted in \autoref{fig:study-setup}.

\textbf{Methodology}
At the beginning of the study, participants were asked to rate their experience in software development, e.g., regarding web development.
Then, they use of the live tracing capabilities of ExplorViz to inspect an instrumented variant of the Spring PetClinic sample application (PetClinic).\footnote{\url{https://github.com/spring-projects/spring-petclinic}}
First, participants are asked to open the PetClinic in a separate browser and interact with it. Then, they are asked to examine the ExplorViz visualization and answer questions about the occurrence of method calls.
Secondly, an analogous task is performed using the embedded browser.
After this section of the survey, the participants were asked to decide which  way of interacting with the PetClinic they preferred.

The second section of the tasks in our evaluation focused on the HTML inspection and exploration.
The participants were first introduced to the elements inspector of the Google Chrome DevTools.\footnote{\url{https://developer.chrome.com/docs/devtools}}
With this knowledge, they were asked to inspect and determine the type and composition of different HTML elements.
Then again, analogous tasks were asked to be performed using the embedded browser and HTML visualization.
This step of the study is depicted on the large screen in \autoref{fig:study-setup}.

\begin{figure}[htbp]
	\includegraphics[width=\textwidth/2]{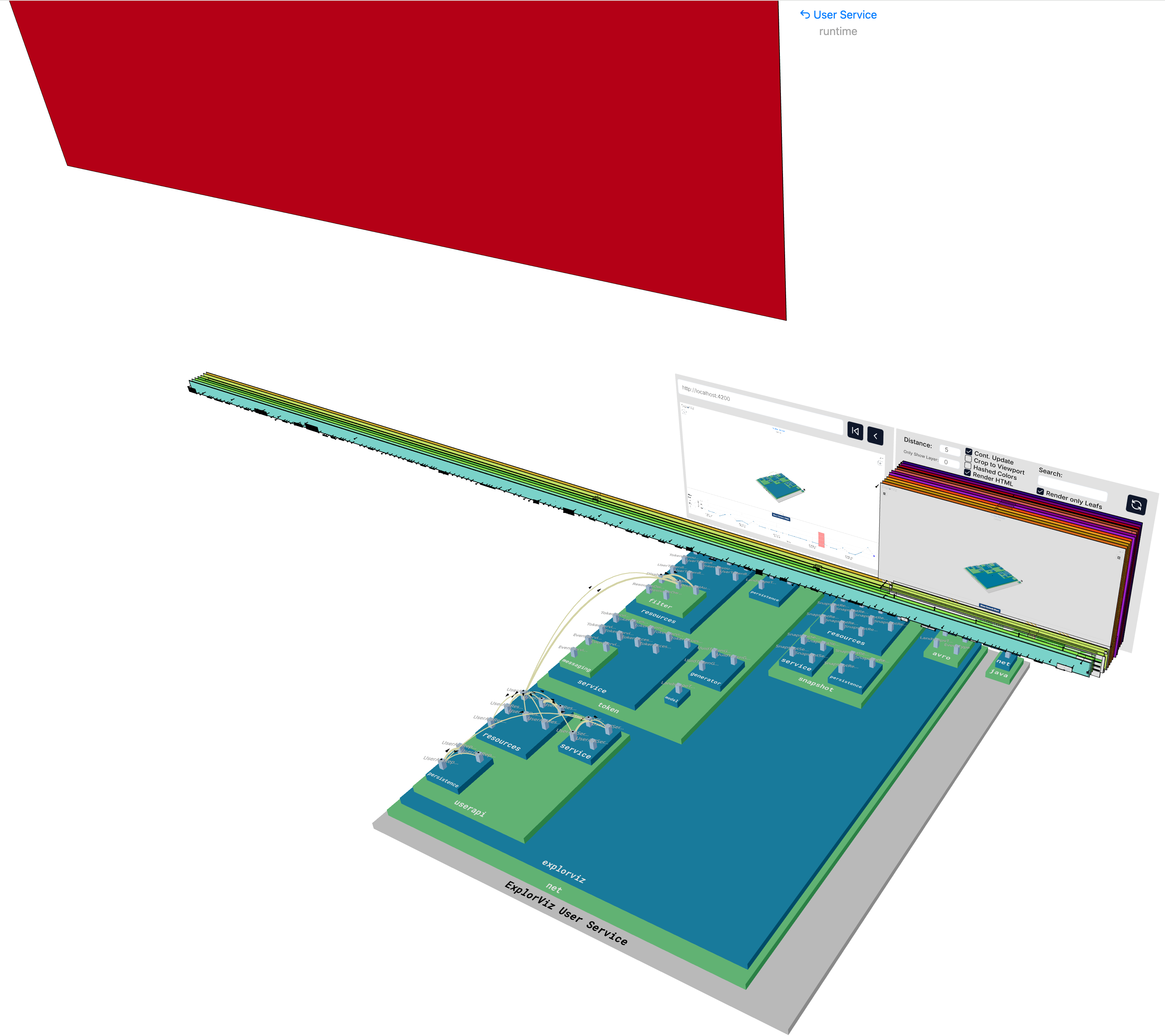}
	\caption{Analysis of the ExplorViz front-end. By disabling viewport cropping, the actual width of the timeline and a large HTML element by plotly.js (in red) is visualized.}
	\label{fig:study-viewport}
\end{figure}

The last task asked the participants to explore the main visualization of ExplorViz.
The participants were asked to turn off the viewport cropping option, such that overflowing (scrollable) content and any other HTML elements outside the browser's viewport are displayed.
The participants were then asked to name any unexpected elements outside the viewport.
It was intended that participants find a large svg-element outside the viewport, which is generated by plotly.js\footnote{\url{https://github.com/plotly/plotly.js}} for testing purposes as displayed in \autoref{fig:study-viewport}.

At the end of the questionnaire, the study participants were asked to rate their experience and give textual feedback.

\textbf{Results}
Regarding the embedded web view, 4 out of 6 participants preferred the embedded web view over a dedicated browser tab for the given task.
3 participants strongly agree with the statement that the web view is a useful addition, while 2 participants agree and one participant disagrees.
Positive textual feedback includes that the participants do not need to know or keep track of web addresses to use the web interface of an application and can thus quickly produce trace data.
Negative feedback mentions the small size of the embedded web view and states that orienting the camera appropriately can be cumbersome.

Regarding the HTML visualization, 5 participants agree and 1 participant strongly agrees that the visualization helps to understand the structure of the DOM.
Exploration of the DOM with the HTML visualization was perceived as enjoyable by 5 participants (3 agree, 2 strongly agree) while 1 participant rates this as neutral.
In their textual feedback, the participants positively mention that the nesting of HTML elements becomes clear.
The search, filter, and subtree selection options are well perceived.
The option to apply textures to the boxes was also positively mentioned.

Negative feedback for the HTML visualization is mostly concerned with the camera controls, popovers which occlude the view, perceived performance issues, and bugs in the implementation.
The participants provided suggestions for more filter options, visual decluttering, improved search and element counter, as well as ways to solve issues regarding size and positioning of the HTML visualization.

We provide the survey including all answers of the participants in a publicly available package~\cite{hansen_2025_15518604}.

\textbf{Discussion}
The embedded web view received mixed feedback as its size and navigation could be cumbersome.
Looking at the study results, we see the strengths of the embedded web view in the convenience to open the correct website with a single click.
This may be especially helpful in software landscapes with multiple applications.
Without an option to enter full-screen mode in the web view (thereby acting like a separate browser), we conclude regarding our first research question that the embedded browser may only be a viable alternative for short-term interactions with a web application.
More exhaustive exploration of a web interface will most likely be done in a separate browser tab.

Regarding the HTML visualization, the gathered feedback indicates that the 3D visualization of the DOM and HTML elements is easy to understand and thus may also help users understand the underlying HTML structure.
Also, the display of elements outside the viewport offers new perspectives on the given website.
Regarding our second research question, we do not see our HTML visualization as a full replacement for established web inspection tools.
Those are more powerful in terms of in-depth analysis and DOM manipulation.
However, our visualization approach is very accessible and may be used for exploratory tasks.
A possible use case could be the introduction of new developers to web development projects.
Another possible application is educating pupils or students about web interfaces.

\textbf{Threats to Validity}
Regarding the hardware, no well-funded statements can be made about the applicability of our approach for other device configurations. 
Due to the small number of participants (six), the results of the quantitative study are not suitable for generalization.
In addition, all study participants knew the conductor of the study personally and were involved in the development of ExplorViz.
This might induce a bias in favor of our implemented approach.
Although the study participants have acquired experience in web development, they are not representatives of the group of professional front-end developers.

\section{Conclusions and Future Work}\label{sec:conclusion}
In this paper we presented an approach which extended a software city visualization with an embedded web view and an accompanying HTML structure visualization.
The approach is made concrete by means of an open source implementation in our software visualization tool ExplorViz.
A preliminary study has been conducted to demonstrate the applicability of our approach and gather valuable feedback for future research directions.

Beyond the future work implied by the user study, we plan to extend our approach to support live-editing HTML in code editors.
This would enable the combined evaluation of made changes in terms of a website's design and functional behavior through the software city visualization. 
In addition, the use of our approach in virtual reality (VR) environments~\cite{VISSOFT2015VR} is promising, as the availability of web inspection tools is limited for VR.

\section*{Acknowledgments}
We want to thank all participants of our user study for their time and valuable feedback.

\providecommand{\doi}[1]{DOI: \href{https://doi.org/#1}{#1}}
\bibliographystyle{myIEEEtran}
\bibliography{bibliography}

\end{document}